\documentclass[12pt]{article}
\usepackage{amsmath,amsfonts,amssymb,graphicx,epsfig}
\usepackage{graphicx}

\begin{document}
\title{\Huge\bf The role of observers in the measurement of the Teleparallel Gravitoelectromagnetic fields in different geometries}
\author{{{E.P. Spaniol}$^{\,a}$\thanks{E-mail: spaniol@fis.unb.br}, {L.R.A. Belo}$^{\,a}$\thanks{E-mail: leandrobelo@fis.unb.br}, {J.A. de Deus}$^{\,a}$\thanks{E-mail: julianoalves@fis.unb.br}, and {V.C. de Andrade}$^{\,a}$\thanks{E-mail: andrade@fis.unb.br}}\\ \\
\it $^a$ \small \it Physics Institute, Brazilia University\\ \small \it 70.917-910, Bras\'{\i}lia, Federal District, Brazil}

 \date{}
\maketitle
\begin{abstract}
In the context of the Teleparallel Equivalent of General Relativity
(TEGR) we have investigated the role of local observers, associated with
tetrad fields, in description of the gravitational interaction
through the concepts of the gravitoelectric (GE) and
gravitomagnetic (GM) fields. We start by analyzing the
gravitoelectromagnetic (GEM) fields obtained from an observer
freely falling in the Schwarzschild space-time. Then, we
investigated whether it is possible to distinguish between this
situation and to be at rest in the Minkowski space-time. We conclude that, although
there are non-zero components for the fields obtained for the case
of free fall, its dynamical effect, measured by the gravitational
Lorentz force, is null. Moreover, the gravitational field energy
obtained from the GEM fields for an observer freely falling in the
Schwarzschild spacetime is zero. These results are in complete agreement with the equivalence principle.
\textbf{keywords:} Teleparallel Equivalent of General Relativity, Gravitoelectromagnetism, Free falling frame.

\end{abstract}

\newpage

\section{Introduction}
\indent The study of gravitation through the GEM fields can bring some new insight \cite{literatura}. To give a contemporary example, the interpretation of objects recently defined by the literature, as the vortex and tendex lines \cite{owen} can be facilitated if we use the fields to describe them \footnote{This possibility is still under investigation.}. Recently, at the linear regime of gravitation, we had a new confirmation of the existence of the GEM fields, through the Gravit Probe B experiment \cite{gpb}.

In the literature, we found several studies that address the
stationary GEM fields due to the fact that the similarity between
the field equations of electromagnetic theory and of gravitation is
reached in this context \cite{mashhoon}. There are still some
works that deal with the time-dependent GEM and the issue about the
Faraday's law in the context of general relativity (GR)
\cite{mashhoon1}. However, there is no studies about the behavior of GEM fields for cases in which the observer is moving with respect to the source.

On the other hand, with the advent of relativistic mechanics, the
relativity principles were extended to the electrodynamics in
analyzing the behavior of electric and magnetic fields in relation
to different inertial frames. Just as in electrodynamics, we
expect that, in gravitation, the GEM fields proposed assume
different expressions depending on how the source is observed, i.e.,
different observers note different GEM fields. Thus, it is natural to study the physical consequences of these different fields upon the observer itself.

In a previous work \cite{spaniol}, motivated by the fact that the
TEGR can be described as a gauge theory, we have proposed a new way
to define the gravitoelectric and the gravitomagnetic fields.
These definitions, that are conceptually different from those that
arise in the RG, were made in a very similar way to what is done on
the Yang-Mills theory and the Electromagnetism, being based on the
field strength components. On that work, in the weak field limit we have obtained the analogous Maxwell equations and for a set of tetrads which is adapted to a stationary observer relative to Schwarzschild,
the gravitoelectric components calculated were in total agreement with the
newtonian field.

According to \cite{malufquedalivre} we can interpret the extra
degrees of freedom of the tetrad field as a choice of reference
system. Two sets of tetrad fields may represent the same spacetime,
though they are physically different. That is, besides being the
fundamental object of the theory, we can interpret them as ideal
observers in spacetime. This subtleness is not present in the metric
description of gravity.

In this paper, on the context of the TEGR, we discuss the issue of
how different observers feel the GEM fields. For two different
observers, we will analyze the behavior of GEM fields for the
Schwarzschild and the Minkowski spacetime. First, we will consider a
free falling observer in Schwarzschild black hole, ie an observer
who falls radially into a black hole due to its gravitational force,
then we will consider a second observer, but now, the observer is
standing in the Minkowski spacetime. As expected, from the
equivalence principle, we have concluded in this approach that those
observers are indistinguishable from the dynamic point of view, that
is, being null the gravitational Lorentz force \cite{andrade} felt
by each one of them, they follow the same trajectory. Moreover,
their energies were calculated, being zero in both cases. It is
interesting to understand how these results occur even though we
have found non-zero GEM components, as can be viewed along the
subsections 2.1 and 2.2.

Notation: According to its gauge structure, to each point of
spacetime there is attached a tangent spacetime (the fiber of the
correspondent tangent bundle), on which the gauge group acts, and
whose metric is assumed to be $\eta_{ab}=(-1,1,1,1)$. The spacetime
indices will be denoted by the Greek alphabet $(\mu,\nu,\sigma,...$
$= 0,1,2,3)$ and the tangent space indices will be denoted by the
first half of the Latin alphabet $(a,b,c.. = 0,1,2,3)$.  The second
half of the Latin alphabet will be used to represent space tensor
components, that is, $(i,j,k...)$ assume the values 1,2 and 3.
Indices in parentheses will also be related to tangent space. We
adopt the light velocity as $c=1$.

\subsection{Teleparallel Equivalent of General Relativity and Gravitoelectromagnetism}
Let us present some of the more important expressions in TEGR that
will be used in the whole paper \footnote{For detailed of the
teleparallel fundamentals see Ref.\cite{andrade,andrade2}.}.

The field strength of the theory is defined in the usual form
\begin{equation}
F^{a}{}_{\mu \nu }=\partial _{\mu }h^{a}{}_{\nu }-
\partial_{\nu}h^{a}{}_{\mu } = h^{a}{}_{\rho }\;T^{\rho }{}_{\mu \nu } \;,  \label{fb}
\end{equation}
with $h^{a}{}_{\mu }$ being the components of the tetrad field. The
object $T^{\rho }{}_{\mu \nu }$ is the torsion that represents alone
the gravitational field, defined by $T^\rho{}_{\mu \nu}
\;\equiv\Gamma^\rho{}_{\nu \mu} \;-\Gamma^\rho{}_{\mu \nu}$, where
$\Gamma^\rho{}_{\nu \mu}$ is the Weitzenb\"ock connection given by
$\Gamma^\rho{}_{\nu \mu} \;\equiv
h_{a}{}^{\rho}\partial_{\mu}h^{a}{}_{\nu}$. Therefore, torsion can also
be identified as the field strength written on the tetrad base.

The dynamics of the gauge fields will be determined by the lagrangian \cite{malufsuperpotencial}
\begin{equation}
\mathcal{L}_{G}=\frac{h}{16\pi G}\;S^{\rho \mu \nu }\;T_{\rho \mu
\nu }\;,
\end{equation}
with $h=\mathrm{det}(h^{a}{}_{\mu })$ and
\begin{equation}
S^{\rho \mu \nu }=-S^{\rho \nu \mu }\equiv {\frac{1}{2}}\left[
K^{\mu \nu \rho }-g^{\rho \nu }\;T^{\theta \mu }{}_{\theta }+g^{\rho
\mu }\;T^{\theta \nu }{}_{\theta }\right] \label{superpotencial}
\end{equation}
which is called superpotential, that will play an important role in
theory, as we will see. The object $K^{\mu \nu \rho }$ is the contorsion tensor
defined by
\begin{equation}
K^{\mu \nu \rho }= \frac{1}{2} \, T^{\nu \mu \rho}+\frac{1}{2} \, T^{\rho \mu \nu}- \frac{1}{2}T^{\mu \nu \rho}.
\end{equation}

The field equations resulting from this lagrangian are
\begin{equation}
\partial _{\sigma }(hS_{a}{}^{\sigma \rho }) - 4\pi G(hj_{a}{}^{\rho })=0  \label{EQcampo}
\end{equation}
with
\begin{equation}
j_{a}{}^{\rho }\equiv \frac{\partial{\mathcal
L}}{\partial{h^{a}{}_{\rho}}}
=h_{a}{}^{\lambda}(F^{c}{}_{\mu\lambda}S_{c}{}^{\mu\rho} -
\frac{1}{4}
\delta_{\lambda}{}^{\rho}F^{c}{}_{\mu\nu}S_{c}{}^{\mu\nu}),
\label{ptemp2}
\end{equation}
being the gauge energy-momentum current of the gravitational field \cite{andrade2}.

Let us now introduce the gravitoelectromagnetism in the
teleparallel context. On one hand, in the context of TEGR, the field
strength $F^{a}{}_{\mu\nu}$ can be associated to the torsion tensor,
in such way that we could use it to define our fields. On the other
hand, the superpotential, defined above, assumes the role of the
field strength in the field equations, similarly to what occurs in
the electromagnetic equations. Therefore, inspired on the
electromagnetism, we define the gravitoelectric and
gravitomagnetic fields in terms of the superpotential components. The gravitoelectric field (GE) is defined by
\begin{equation}
S_{a}{}^{0 i}\equiv E_{a}{}^{i}
\label{GE}
\end{equation}
and the gravitomagnetic field (GM) is as follows
\begin{equation}
S_{a}{}^{i j}\equiv \epsilon^{i j k}B_{a k}. \label{GM}
\end{equation}
As already stated, these definitions were tested and passed on some important tests \cite{spaniol}. Calculating these fields in specific configurations will allow us to better understand the role of observers in gravitation.

\section{Free Falling in Schwarzschild spacetime}
In this section we analyze the GEM fields obtained from an observer
in free fall in Schwarzschild spacetime. Initially we consider the
Schwarzschild metric which can be written as
\begin{equation}
ds^2 = -\alpha^{-2} dt^2 + \alpha^{2}dr^2 + r^2(d\theta^2 + {sin \theta}^2 d\phi^2) \label{metricaschw}
\end{equation}
with
\begin{equation}
\alpha^{-2} = 1 - \frac{2m}{r}.
\end{equation}

An observer that moves radially in free fall due to
attraction of the Schwarzschild black hole must have a
four-velocity like \cite{hartle}
\begin{equation}
u^{\nu} = \Big [ \Big(1 - \frac{2m}{r} \Big)^{-1}, -
\Big(\frac{2m}{r}\Big)^{1/2} , 0, 0 \Big]. \label{quadrivelocidade}
\end{equation}
A set of tetrad fields that satisfies the above condition is given
by \cite{malufquedalivre}
\begin{equation}
h_{a \mu}=
\left(%
   \begin{array}{cccc}
          -1& -\alpha^2 \beta& 0 & 0 \\
          \beta sin\theta cos\phi & \alpha^2 sin\theta cos\phi & rcos\theta cos\phi & -rsin\theta sin\phi\\
          \beta sin\theta sen\phi & \alpha^2 sin\theta sen\phi & rcos\theta sin\phi & rsin\theta cos\phi \\
          \beta cos\theta& \alpha^2 cos\theta & -rsin\theta & 0\\
   \end{array}%
\right),\label{tetradasch}
\end{equation}
where $\beta$ is defined by
\begin{equation}
\beta =\sqrt{\frac{2m}{r}}.
\end{equation}
Through the expression of torsion written in terms of the tetrad fields
\begin{equation}
T^{\sigma }{}_{\mu \nu }=h_{a}{}^{\sigma }\partial _{\mu }h^{a
}{}_{\nu } - h_{a}{}^{\sigma }\partial _{\nu }h^{a }{}_{\mu
},\label{Thh}
\end{equation}
we calculate the components of $T_{\sigma\mu \nu }$, of which the
non null are
\begin{eqnarray} \nonumber
T_{001} &=& - \beta \partial_r \beta, \\ \nonumber T_{101} &=&
-\alpha^{2} \partial_r \beta, \\ \nonumber T_{202} &=& - r \beta, \\
\nonumber T_{303} &=& - r \beta \sin^2 \theta, \\ \nonumber T_{212}
&=& r (1-\alpha^2), \\ T_{313} &=& r (1-\alpha^2) \sin^2 \theta.
\end{eqnarray}

With these results, we can calculate the superpotential that will
allow us to find the GEM fields. In this way, to get the GE fields,
we need the componentes of $S_b{}^{\mu\nu}$ following:
\begin{equation}
S_b{}^{0i} = \frac{1}{4} \Big[ h_b{}^{k}g^{00}g^{ij}T_{j0k} +
h_b{}^{k}g^{00}g^{ij}T_{k0j} \Big] + \frac{1}{2} \Big[
h_b{}^{0}g^{ij}g^{lk}T_{kjl} - h_b{}^{i}g^{00}g^{kj}T_{j0k} \Big].
\label{GE1}
\end{equation}
The GE radial components are obtained by making $i=1$, that is
\begin{equation}
S_b{}^{01}=E_b{}^{1} = \frac{1}{2} \Big[ h_b{}^{0}g^{11}g^{22}T_{212} +
h_b{}^{0}g^{11}g^{33}T_{313} - h_b{}^{1}g^{00}g^{22}T_{202} -
h_b{}^{1}g^{00}g^{33}T_{303} \Big]. \label{309}
\end{equation}
To the angular components $\theta$ we perform $i=2$ in the above
expression (\ref{GE1}) and we find
\begin{equation}
S_{b}{}^{0 2}=E_b{}^{2}= -\frac{1}{2} \left[ h_b{}^2 g^{00}g^{11}T_{101} +
h_b{}^2 g^{00}g^{33}T_{303} \right].\label{319}
\end{equation}
The $\phi$ components are obtained by making $i=3$
\begin{equation}
S_{b}{}^{0 3}=E_b{}^{3}= -\frac{1}{2} \left[ h_a{}^3 g^{00}g^{11}T_{101} +
h_a{}^3 g^{00}g^{22}T_{202} \right]. \label{329}
\end{equation}
Let us now consider the internal index equal to zero in the above
expressions, that is, $b=0$:
\begin{eqnarray} \nonumber
E_{(0)}{}^{r} = 0, \\ \nonumber E_{(0)}{}^{\theta}=0, \\
E_{(0)}{}^{\phi}=0. \label{ge0}
\end{eqnarray}
Then we calculate the spacial components for $b$. Considering
(\ref{309}) and attributing $b=1,2,3$ we get
\begin{equation}
E_{(1)}{}^{r} = - \frac{\beta}{r}\sin\theta \cos\phi,
\end{equation}
\begin{equation}
E_{(2)}{}^{r} = - \frac{\beta}{r}\sin\theta \sin\phi,
\end{equation}
\begin{equation}
E_{(3)}{}^{r} = - \frac{\beta \cos\theta}{r}.
\end{equation}
In the same way, assigning the values $b=1,2,3$ in (\ref{319}), we
obtain
\begin{equation}
E_{(1)}{}^{\theta}= - \frac{\alpha^2\beta}{4r^2}\cos\theta\cos\phi,
\end{equation}
\begin{equation}
E_{(2)}{}^{\theta}= - \frac{\alpha^2\beta}{4r^2}\cos\theta\sin\phi,
\end{equation}
\begin{equation}
E_{(3)}{}^{\theta}= - \frac{\alpha^2\beta}{4r^2}\sin\theta,
\end{equation}
and finally, making $b=1,2,3$ in (\ref{329}) we get
\begin{equation}
E_{(1)}{}^{\phi}= \frac{\alpha^2\beta \sin\phi}{4r^2 \sin\theta},
\end{equation}
\begin{equation}
E_{(2)}{}^{\phi}= - \frac{\alpha^2\beta \cos\phi}{4r^2 \sin\theta},
\end{equation}
\begin{equation}
E_{(3)}{}^{\phi}=0 .
\end{equation}

Let us now calculate the GM fields for this configuration. Writing
the superpotencial in terms of torsions,
\begin{eqnarray}\nonumber
S_b{}^{ij}&=& \frac{1}{4} \left[ h_a{}^{0}g^{ik}g^{jm} \left(
T_{mk0} + T_{0km} - T_{km0} \right) + h_a{}^{n}g^{ik}g^{jm} \left(
T_{mkn} + T_{nkm} - T_{kmn} \right) \right] \\ &+&  \frac{1}{2}
\left[ - h_a{}^{j}g^{ik}( g^{nm}T_{mkn} -g^{00}T_{00k})
+h_a{}^{i}g^{jl}(g^{nm}T_{mln} - g^{00}T_{00l} ) \right],
\label{GM1}
\end{eqnarray}
and using the definition (\ref{GM}) with the internal index $b=0$ in
the above expression we obtain:
\begin{eqnarray} \nonumber
B_{(0) \phi} = 0, \\ \nonumber B_{(0) \theta} = 0, \\ B_{(0) r} = 0.
\end{eqnarray}
In the sequence we consider $b=1,2,3$ for each spacetime coordinate.
For $\phi$ component:
\begin{equation}
B_{(1) \phi} =  \frac{m}{2 r^3}\cos\theta\cos\phi,
\end{equation}
\begin{equation}
B_{(2) \phi} =  \frac{m}{2r^3}\cos\theta\sin\phi,
\end{equation}
\begin{equation}
B_{(3) \phi} = - \frac{m}{2r^3}\sin\theta.
\end{equation}
For $\theta$ component:
\begin{equation}
B_{(1) \theta} =  \frac{m}{2r^3} \frac{\sin\phi}{\sin\theta},
\end{equation}
\begin{equation}
B_{(2) \theta} = - \frac{m}{2r^3} \frac{\cos\phi}{\sin\theta},
\end{equation}
\begin{equation}
B_{(3) \theta} = 0.
\end{equation}
Finally, the remaining radial components
\begin{equation}
B_{(1) r}=B_{(2) r}=B_{(3) r}=0.
\end{equation}

On the other hand, if we consider a static observer in Minkowski
spacetime and perform a similar calculation we obtain all the GEM
field components equal to zero. However, assuming valid the equivalence
principle, we should not be able to discern between two observers,
one of then freely falling in Schwarzschild black hole, and other
static in Minkowski spacetime. This apparent inconsistency should be
clarified when investigating the role of the non-zero components $b
= 1,2,3$ for dynamics.

Before doing this analysis, let us make some comments. According to
\cite{Schmid} in the linearized GEM the operational \footnote{That
allow a direct analogy with electromagnetism.} definition for the
GEM fields must be in accordance with the equivalence principle,
that is, for a non-rotating and free fall observer there is no
gravitational forces and therefore the GEM fields are zero. Our
definition is in full agreement with this since in the weak field
limit
\begin{equation}
\frac{m}{r}<<1
\end{equation}
all the above components are zero. Moreover, even in the exact case,
we have shown that the $b=0$ components also vanish. This show that
the operational definitions must be related with $b=0$ components, which is in agreement with a similar analysis in \cite{spaniol}.

Let us now verify the effects of the non-zero components of the GE
and GM fields on the dynamics of the observers.

\subsection{Gravitational Lorentz Force}
As mentioned earlier, as a consequence of the equivalence principle,
an observer represented by a not spinning tetrad field and freely
falling in Schwarzschild spacetime, should not be able to
distinguish - at least from the dynamic point of view - if is freely
falling in this spacetime or at rest with respect to the Minkowski
spacetime. A way to tackle this issue is to use the equation that
describes the behavior of scalar particles in the presence of
gravitation: the TEGR gravitational Lorentz force \cite{andrade}
\begin{equation}
h^{a}{}_{\mu}\frac{du_a}{ds}=F^{a}{}_{\mu\nu}u_{a} u^{\nu}, \label{lore1}
\end{equation}
in which the right side of the equation plays the role of force,
analogous to the Lorentz force of Electromagnetism. Alternatively,
this equation can be rewritten as the geodesic equation in the
context of RG \cite{andrade}. From this expression, we can evaluate
the consequences of the non-zero GEM fields components previously
obtained.

Since the GEM fields are defined from the superpotencial $S^{b \rho
\mu }$ it is convenient to rewrite the above equation in terms of
these quantities. For this, we should first rewrite the
gravitational field strength tensor in terms of the superpotential,
ie
\begin{equation}
F^{a}{}_{\gamma \delta}=h^{b}{}_{\gamma}g_{\rho
\delta}h^{a}{}_{\mu}S_{b}{}^{\mu \rho}- h^{b}{}_{\delta}g_{\nu
\gamma}h^{a}{}_{\mu}S_{b}{}^{\mu
\nu}-\frac{1}{2}h^{a}{}_{\delta}g_{\nu
\gamma}h^{b}{}_{\theta}S_{b}{}^{\theta\nu}+\frac{1}{2}h^{a}{}_{\gamma}g_{\rho
\delta}h^{b}{}_{\theta }S_{b}{}^{\theta \rho}.\label{fs}
\end{equation}
Thus, we obtain
\begin{equation}
h^{a \mu} \frac{d u_a}{d s} = - h_{b}{}^{\nu}S^{b \rho
\mu}u_{\rho}u_{\nu} -\frac{1}{2}h_{b \theta}(S^{b \mu
\theta}u^{\rho}u_{\rho} - S^{b \nu \theta}u^{\mu}u_{\nu}).
\label{lorentz2}
\end{equation}

By making use of the (\ref{quadrivelocidade}) and of the GEM fields
we can calculate the right side of the equation (\ref{lorentz2}) for
an observer freely falling in Schwarzschild spacetime. Thus, we get
a null result for all Lorentz force components, i.e., the non-null GEM
fields, obtained in earlier section, are combined so as to eliminate
the force felt by the observer, and therefore do not changing its
trajectory. The same result is obtained when we consider a static
observer in relation to the Minkowski spacetime, since all GEM
fields are null. Therefore, in some sense, we can say that the
components of superpotencial with zero internal space index
represent the operational definition of the GEM fields since, being
equal to zero, these components were already in line with the
equivalence principle.

\subsection{Gravitational Field Energy}
Another physical evidence that enables us to face the issue of
non-zero components for the case of the freely falling reference
frame in the Schwarzschild spacetime is the gravitational field
energy. Again, being valid the equivalence principle, we should not
to be able to discern between two observers, one freely falling in
Schwarzschild black hole, and another static in Minkowski spacetime.
Thus, being zero the gravitational field energy associated with the
second situation, an equal result should occur with the energy
measured by the observer in the first situation. We can calculate
the gravitational field energy as given by the zero component of
(\ref{ptemp2}), from the GEM fields obtained by an observer
represented by the tetrad field (\ref{tetradasch}), since they are
defined from the superpotential which appears in the definition of
energy momentum tensor. Let us consider then
\begin{equation}
j_{(0)}{}^{0}=h_{(0)}{}^{\lambda}(F^{c}{}_{i\lambda}S_{c}{}^{i 0} -
\frac{1}{4}
\delta_{\lambda}{}^{0}F^{c}{}_{\mu\nu}S_{c}{}^{\mu\nu}). \label{c00}
\end{equation}
Substituting (\ref{fs}) in (\ref{c00}) we get
\begin{eqnarray} \nonumber
j_{(0)}{}^{0} &=& h_{(0)}{}^{\lambda}\Big(h^{b}{}_{i}
g_{\rho\lambda}h^{c}{}_{\gamma}S_{b}{}^{\gamma\rho} -
h^{b}{}_{\lambda} g_{\rho i}h^{c}{}_{\gamma}S_{b}{}^{\gamma\rho} -
\frac{1}{2}h^{c}{}_{\lambda} g_{\rho
i}h^{b}{}_{\gamma}S_{b}{}^{\gamma\rho} \\ \nonumber
&+& \frac{1}{2}h^{c}{}_{i}
g_{\rho \lambda}h^{b}{}_{\gamma}S_{b}{}^{\gamma\rho} \Big)S_{c}{}^{0
i} + \frac{1}{4}h_{(0)}{}^{0} \Big(h^{b}{}_{\mu}
g_{\rho\lambda}h^{c}{}_{\gamma}S_{b}{}^{\gamma\rho} -
h^{b}{}_{\lambda} g_{\rho\mu}h^{c}{}_{\gamma}S_{b}{}^{\gamma\rho} \\
&-&\frac{1}{2}h^{c}{}_{\lambda}
g_{\rho\mu}h^{c}{}_{\gamma}S_{b}{}^{\gamma\rho}+\frac{1}{2}h^{c}{}_{\mu}
g_{\rho\lambda}h^{b}{}_{\gamma}S_{b}{}^{\gamma\rho} \Big)
S_{c}{}^{\mu\lambda}. \label{energiaquadratica}
\end{eqnarray}
Using the definitions (\ref{GE}) and (\ref{GM}) we can rewrite the
expression above in terms of $E_{a}{}^{i}$ and $B_{a}{}^{i}$. Note
that as it is quadratic in the superpotential, it is also quadratic
in the GEM fields. After a lengthy calculation, we found out the
following result for the above component
\begin{equation}
j_{(0)}{}^{0}=0, \label{energiazero}
\end{equation}
ie the gravitational field energy written in terms of the GEM fields
are zero for a freely falling observer in the Schwarzschild black
hole. While there are non-zero components of the GEM field, they
combine in such a way that do not change the expected result of the
gravitational field energy, in a similar way with what happened in
gravitational Lorentz force calculation. As consequence, it is not
possible - at least from the dynamical point of
view\footnote{Perhaps the components $E_{(1,2,3)}{}^{i}$ and
$B_{(1,2,3)}{}_{k}$ have a measurable physical sense in a semi-
classical scenario and allow a differentiation between the frames.
The analysis of this issue will be presented elsewhere.} - for a
local observer to distinguish  between to be in free fall in the
Schwarzschild black hole or to be at rest in the Minkowski
spacetime.

Thus, from (\ref{energiaquadratica}) and (\ref{energiazero}), we can define an "operational energy" of the gravitational field in a manner completely analogous to that of electromagnetism, namely:
\begin{equation}
P=\int \left[\left(E_{(0)}{}^{i}\right)^2+\left(B_{(0)i}\right)^2\right]d^3x.
\end{equation}
It is important to stress out that this definition was inferred based only on the case of a reference in free fall in a Schwarzschild black hole, being the extension of its validity still under investigation.

\section{Final remarks}

As obtained in a previous work \cite{spaniol}, for a set of tetrads
which is adapted to a stationary observer relative to Schwarzschild
spacetime, it has been showed that in the weak field limit the
gravitoelectric components are in total agreement with the
Newtonian field and, in addition, all GM components are zero. The
conceptual definitions of what we expect to be analogous to
electromagnetic fields were identified as the zero internal
components of GEM fields, that is $b=0$, what we have called
"operational definitions". Here in this work, when we consider a freely
falling observer in the Schwarzschild black hole we obtained a null
result for all the GE and GM components with zero internal index,
which corroborate the idea of "operationality" for $b=0$ component
fields. We would like to emphasize that the choice of coordinate
systems is the same in both cases above mentioned, through the use
of appropriate tetrad fields.

We have obtained as main conclusion in this work that through the
use of GEM fields it is not possible for a local observer to
distinguish between free falling in the Schwarzschild black hole or
at resting in the Minkowski spacetime. Although this idea seems to
be natural, due to the equivalence principle, it emerged in this
approach after a deeper analysis, since we found out that non-null
fields arise in the free fall case. One possibility would be to
consider only the operational components $b=0$, since they are all
equal to zero and simply to discard the other non null components
that came from spacial internal indices. Then it would be
straightforward to postulate the equivalence between the references.
But these no null components could store some important information
that would violate the central idea.

To investigate the role of non null components of GEM fields in
dynamics we have used the gravitational Lorentz force written in terms of
them and we concluded that their contributions cancel each other
resulting in a null total force measured by the free falling
observer in the Schwarzschild geometry, in the same way it were
placed at rest in Minkowski spacetime . Thus, any experiment which
make use of dynamical effects from the gravitational field will not
be able to distinguish between those two references. Moreover, in
order to support the results, we showed that the gravitational field
energy measured by the reference in free fall is zero, as expected
if compared with the field energy associated with the flat
spacetime. We also should like to stress out that all the
calculations were done using the GEM fields and outside the weak
field limit, that is, we have obtained exact results that can also be
applicable to treat intense fields like, for example, jet formations in supermassive black holes.
\vskip .8cm
\centerline{\bf \Large Acknowledgements\\}
\vskip .5cm \noindent The authors thank CAPES and CNPq (Brazilian agencies) for the financial support.\\

\end{document}